\documentclass[a4paper,fleqn,usenatbib]{mnras}

\usepackage{mathtools}  
\usepackage{amssymb}
\usepackage[T1]{fontenc}
\usepackage{ae,aecompl}
\usepackage{adjustbox}

\usepackage{lipsum}

\usepackage{graphicx}	
\usepackage{amsmath}	
\usepackage{amssymb}	
\usepackage{natbib} 

\usepackage{xcolor}
\definecolor{bluegray}{rgb}{0.20, 0.60, 0.80}

\title[Zeeman effect in the solar Mg~{\sc i}~$b$ lines]{Study of the polarization produced by the Zeeman effect in the solar Mg~{\sc i}~$b$ lines}

\author[C. Quintero Noda et al.]{C. Quintero Noda,$^{1}$\thanks{E-mail: carlos@solar.isas.jaxa.jp}
H. Uitenbroek,$^{2}$
M. Carlsson,$^{3,4}$
D. Orozco Su\'arez,$^{5}$
\newauthor
Y. Katsukawa,$^{6}$
T. Shimizu,$^{1}$
B. Ruiz Cobo,$^{7,8}$
M. Kubo,$^{6}$
T. Oba,$^{1}$
\newauthor
Y. Kawabata,$^{1,9}$
T. Hasegawa,$^{1,9}$
K. Ichimoto,$^{6,10}$
T. Anan,$^{11}$
Y. Suematsu$^{6}$ 
\\
$^{1}$Institute of Space and Astronautical Science, Japan Aerospace Exploration Agency, Sagamihara, Kanagawa 252-5210, Japan\\
$^{2}$National Solar Observatory, University of Colorado Boulder, 3665 Discovery Drive, Boulder, CO 80303, USA\\
$^{3}$Rosseland Centre for Solar Physics, University of Oslo, P.O. Box 1029 Blindern, N-0315 Oslo, Norway\\
$^{4}$Institute of Theoretical Astrophysics, University of Oslo, P.O. Box 1029 Blindern, N-0315 Oslo, Norway\\
$^{5}$Instituto de Astrof\'isica de Andaluc\'ia (CSIC), Glorieta de la Astronom\'ia, 18008 Granada, Spain\\
$^{6}$National Astronomical Observatory of Japan, 2-21-1 Osawa, Mitaka, Tokyo 181-8588, Japan\\
$^{7}$Instituto de Astrof\'isica de Canarias, E-38200, La Laguna, Tenerife, Spain.\\
$^{8}$Departamento de Astrof\'isica, Univ. de La Laguna, La Laguna, Tenerife, E-38205, Spain\\
$^{9}$Department of Earth and Planetary Science, The University of Tokyo, 7-3-1 Hongo, Bunkyo-ku, Tokyo 113-0033, Japan\\
$^{10}$Kwasan and Hida Observatories, Kyoto University, Kurabashira Kamitakara-cho, Takayama-city, 506-1314 Gifu, Japan\\
$^{11}$National Solar Observatory, 22 Ohi'a Ku, Makawao, HI 96768, USA\\
}

\date{Accepted XXX. Received YYY; in original form ZZZ}

\pubyear{2018}

\begin{document}
\label{firstpage}
\pagerange{\pageref{firstpage}--\pageref{lastpage}}
\maketitle

\begin{abstract}
The next generation of solar observatories aim to understand the magnetism of the solar chromosphere. Therefore, it is crucial to understand the polarimetric signatures of chromospheric spectral lines. For this purpose, we here examine the suitability of the three Fraunhofer Mg~{\sc i}~$b_1$, $b_2$, and $b_4$ lines at 5183.6, 5172.7, and 5167.3~\AA, respectively. We start by describing a simplified atomic model of only 6 levels and 3 line transitions for computing the atomic populations of the 3p-4s (multiplet number 2) levels involved in the Mg~{\sc i}~$b$ line transitions assuming non-local thermodynamic conditions and considering only the Zeeman effect using the field-free approximation. We test this simplified atom against more complex ones finding that, although there are differences in the computed profiles, they are small compared with the advantages provided by the simple atom in terms of speed and robustness. After comparing the three Mg~{\sc i} lines, we conclude that the most capable one is the $b_2$ line as $b_1$ forms at similar heights and always shows weaker polarization signals while $b_4$ is severely blended with photospheric lines. We also compare Mg~{\sc i}~$b_2$ with the K~{\sc i}~$D_1$ and Ca~{\sc ii} 8542~\AA \ lines finding that the former is sensitive to the atmospheric parameters at heights that are in between those covered by the latter two lines. This makes Mg~{\sc i}~$b_2$ an excellent candidate for future multi-line observations that aim to seamlessly infer the thermal and magnetic properties of different features in the lower solar atmosphere. 
\end{abstract}



\section{Introduction}

\begin{figure*}
\begin{center} 
 \includegraphics[trim=0 0 0 0,width=15.5cm]{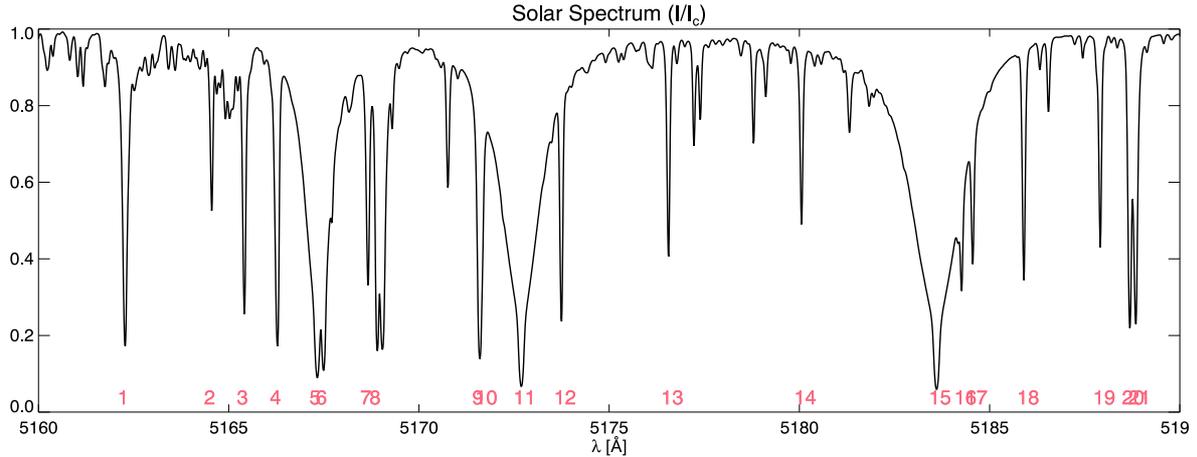}
 \vspace{-0.15cm}
 \caption{Solar atlas at 517~nm extracted from \citet{Delbouille1973}. We label with numbers the lines that are described in Table~\ref{lines}.}
 \label{atlas}
 \end{center}
\end{figure*}

Routine polarimetric observations of spectral lines that form in the chromosphere are expected in the near future. Large ground-based telescopes such as  DKIST \citep{Keil2011} and EST \citep{Collados2013} will have access to various instruments that cover the solar spectrum from approximately 380~nm  to the infrared allowing to perform simultaneous polarimetric observations of multiple solar spectral lines. Moreover, we also have the Sunrise solar balloon-borne observatory \citep{Barthol2011,Berkefeld2011,Gandorfer2011}, which has had two successful science flights in 2009 and 2013 \citep{Solanki2010,Solanki2017} and is preparing a third flight with the aim of observing multiple spectral regions from the near ultraviolet to the near infrared with a set of three instruments. Among them, we have the Sunrise Infrared Spectro-polarimeter (SCIP, Katsukawa et al. in preparation) that will perform spectropolarimetric observations of the 770 and 850~nm windows studied in \cite{QuinteroNoda2017c,QuinteroNoda2017,QuinteroNoda2017b,QuinteroNoda2018} or an updated version of the Imaging Magnetograph eXperiment \citep[IMaX,][]{MartinezPillet2011} renamed as IMaX+. The latter, besides the expected technical improvements that will be described in future publications, will expand the science capabilities of its predecessor (IMaX scanned one spectral line at 525.2~nm) with a second spectral window at 517~nm. The aim is to perform high cadence (less than one minute for the two spectral lines) polarimetric observations of the visible Mg~{\sc i}~$b$ lines, known to be sensitive to higher atmospheric layers \citep[e.g.,][]{Plaskett1931,Waddell1963} than those covered by the photospheric lines at 525.2~nm. Following the strategy used in previous works, we aim to characterize the Mg~{\sc i}~$b$ lines, in order to understand the capabilities and limitations of these spectral lines for inferring the thermodynamic and magnetic properties of the solar atmosphere.

\begin{table}
\hspace{-0.5cm}
\normalsize
\begin{adjustbox}{width=0.450\textwidth}
  \bgroup
\def\arraystretch{1.25}
\begin{tabular}{lccccccc}
	\hline
 & 	Atom                   & $\lambda$ [\AA] & $\log gf$ & $L_l$        & $U_l$            & $\bar{g}$  \\
	\hline
1      & Fe~{\sc i}             &  5162.27      & 0.020  & ${}^5F^{o}_{5}$      &  ${}^5F_{5}$      & 1.40    \\ 
2      & Fe~{\sc i}             &  5164.55     & -1.360  & ${}^3G^{o}_{4}$      &  ${}^3F_{3}$      & 1.00    \\ 
3      & Fe~{\sc i}             &  5165.41     & -0.026  & ${}^5F^{o}_{4}$      &  ${}^5F_{4}$      & 1.35    \\ 
4      & Fe~{\sc i}             &  5166.28     & -4.195  & ${}^5D_{4}$      &  ${}^7D^{o}_{5}$      & 1.80    \\ 
5      & Mg~{\sc i}            &  5167.32     & -0.870  & ${}^3P^{o}_{0}$  &  ${}^3S_{1}$          & 2.00     \\
6      & Fe~{\sc i}             &  5167.48     & -1.260  & ${}^3F_{4}$      &  ${}^3D^{o}_{3}$      & 1.13    \\ 
7      & Ni~{\sc i}              &  5168.65     & -0.430  & ${}^5F^{o}_{3}$      &  ${}^3G_{4}$      & 0.75    \\ 
8      & Fe~{\sc i}             &  5168.89     & -3.969  & ${}^5D_{3}$      &  ${}^7D^{o}_{3}$      & 1.50    \\ 
9      & Fe~{\sc i}             &  5171.60      & -1.793  & ${}^3F_{4}$      &  ${}^3F^{o}_{4}$      & 1.25    \\ 
10    & Fe~{\sc i}             &  5171.67      & -1.912  & ${}^3D_{3}$      &  ${}^1G^{o}_{4}$      & 0.50    \\ 
11    & Mg~{\sc i}            &  5172.68      & -0.393  & ${}^3P^{o}_{1}$  &  ${}^3S_{1}$          & 1.75     \\	
12    & Ti~{\sc i}             &  5173.74      & -1.118   & ${}^3F_{2}$      &  ${}^3F^{o}_{2}$      & 0.67    \\
13    & Ni~{\sc i}             &  5176.55     & -0.440  & ${}^1D^{o}_{2}$      &  ${}^1D_{2}$      & 1.00    \\
14    & Fe~{\sc i}            &  5180.06     & -1.260  & ${}^3G^{o}_{3}$      &  ${}^3F_{2}$      & 0.83    \\
15    & Mg~{\sc i}            &  5183.60      & -0.167  & ${}^3P^{o}_{2}$  &  ${}^3S_{1}$          & 1.25     \\
16    & Fe~{\sc i}            &  5184.27     & -1.000  & ${}^5F^{o}_{2}$      &  ${}^5F_{3}$      & 1.50   \\
17    & Ni~{\sc i}            &  5184.56     & -0.833  & ${}^3D^{o}_{2}$      &  ${}^3P_{1}$      & 1.00    \\
18    & Ti~{\sc ii}           &  5185.91     & -1.350  & ${}^2G_{7/2}$      &  ${}^2G^{o}_{7/2}$      & 0.89    \\
19    & Fe~{\sc i}           &  5187.92     & -1.260  & ${}^3F_{3}$      &  ${}^3D^{o}_{2}$      & 1.00   \\
20    & Ti~{\sc ii}           &  5188.68     & -1.260  & ${}^2D_{5/2}$      &  ${}^2D^{o}_{5/2}$      & 1.25    \\
21    & Ca~{\sc i}           &  5188.84     & -0.090  & ${}^1P^{o}_{1}$      &  ${}^1D_{2}$      & 1.00    \\
	\hline
  \end{tabular}
  \egroup
\end{adjustbox}
\caption {Spectral lines included in the 517~nm window presented in Figure~\ref{atlas}. Each column, from left to right, contains the number assigned to each line, the corresponding atomic species, the line core wavelength, $\log gf$ of the transition, and the spectroscopic notation of the lower and the upper level  \citep[retrieved from the database of R.~Kurucz,][]{Kurucz1995}. The last column contains the effective Land\'{e} factor computed assuming L-S coupling.} \label{lines}     
\end{table}

This work is focused on several main topics. The first one is to develop a simplified Mg~{\sc i}~$b$ model atom that we employ in this publication but that we also aim to use in future numerical studies that are computationally demanding such as 3D synthesis or inversions of the Stokes spectra. Using this simplified atom we thoroughly examine the Mg~{\sc i}~$b$ lines, studying and comparing their properties. This means that we estimate the region of formation of the spectral lines, their sensitivity to different atmospheric parameters, and the maximum polarization signals that we can expect for different magnetic field configurations. Finally, making use of the results from previous works \citep[e.g.,][]{QuinteroNoda2017, QuinteroNoda2017b} we compare the Mg~{\sc i}~$b$ lines with the K~{\sc i} $D$ and Ca~{\sc ii} infrared spectral lines to understand the benefits of observing them simultaneously.

\section{Methodology}\label{method}

\subsection{Spectral lines}

There are three Fraunhofer $b$ lines associated with Mg~{\sc i} transitions in the solar visible spectrum. They are located at 5183.60~\AA \ ($b_1$), 5172.68~\AA \ ($b_2$), and 5167.32~\AA \ ($b_4$) (see labels 5, 11, and 15 in Figure~\ref{atlas} and Table~\ref{lines}). The Fraunhofer $b_3$ line is the Fe~{\sc i} transition located at 5168.91~\AA \ (label 8). Of the three Mg~{\sc i} lines, the one that has been mostly observed in the past is the Mg~{\sc i}~$b_2$ line. That spectral line has been used for solar observations for almost a century, first with photographic techniques, e.g. \cite{Plaskett1931}, and later with photoelectric observations as in \cite{Waddell1963}. In the decade of the 80s, spectropolarimetric observations of the line started with \cite{Stenflo1984} observing plage and network regions and \cite{Lites1988} scanning sunspots. This was later expanded to observations of the scattering polarization produced by the spectral line when pointing to the solar limb \citep{Stenflo1997,Stenflo2000}. A summary of the maximum polarization signals produced by the Mg~{\sc i}~$b_2$ line for the mentioned observations is presented in Table~\ref{Signals}. 
 
We believe that in the cases where $\mu{\sim}1$, the main physical mechanism responsible for the polarization signals is the Zeeman effect as the mentioned publications analysed large concentrations of magnetic field, i.e. network, plage, and sunspot regions. However, in the case of the quiet Sun regions observed at $\mu{\sim}0.1$, the detected polarization signals are produced by scattering processes.

Examining Table~\ref{Signals}, we can see that the amplitude of the polarization signals is low to moderate in comparison with traditional photospheric lines, e.g., Fe~{\sc i} 5250.2~\AA \ or 6302.5~\AA, whose amplitude are sometimes twice the values presented in the mentioned table \citep[for instance, see the reviews of][]{Solanki1993,Borrero2011}. Moreover, additional lines that are sensitive to the physical quantities at similar atmospheric layers, as the Na~{\sc i} $\rm D_1$ \& $\rm D_2$, also produce larger polarization signals \citep[e.g.,][]{Stenflo1984,Stenflo2000} in quiet Sun observations at various heliocentric angles. These signals are produced from both the Zeeman effect and scattering processes. Concerning the Zeeman effect, the main reason is that Mg~{\sc i}~$b_2$ is much wider than the mentioned lines, i.e. Zeeman polarization signals in the weak field regime are proportional to the ratio of the Zeeman splitting over the Doppler width \citep[see, for example, Chapter 9 in the monograph of][]{Landi2004}. We can find a similar example, for instance, for the Ca~{\sc ii} infrared triplet \citep[see the observations presented in][]{Cauzzi2008}.

\begin{table}
\hspace{-0.55cm}
\normalsize
\begin{adjustbox}{width=0.499\textwidth}
  \bgroup
\def\arraystretch{1.25}
\begin{tabular}{lccccccc}
	\hline
Solar region & $\mu$  &	$LP/I_{c}^{QS}$ [\%] & $V/I_{c}^{QS}$ [\%]   & Reference   \\
	\hline
Umbra         & 0.90        &  0.10  &  1.00  &  \cite{Lites1988}    \\
Penumbra      & 0.90        &  2.50  &  10.0  &  \cite{Lites1988}    \\	
Pore          & 0.96        &  0.30  &  3.00  &  \cite{Deng2010}     \\
MPIL          & 0.96        &  0.75  &  1.50  &  \cite{Deng2010}     \\  
Strong Plage  & 0.92        &  ---   &  1.10  &  \cite{Stenflo1984}  \\
Weak Plage    & 0.92        &  ---   &  0.33  &  \cite{Stenflo1984}  \\
Quiet Sun     & 0.10        &  0.16$^{*}$  &  ---   &  \cite{Stenflo2000}  \\
	\hline
  \end{tabular}
  \egroup
\end{adjustbox}
\caption {Polarization signals (approximated values) generated by the Mg~{\sc i}~$b_2$ line for different solar observations and normalized to the continuum intensity. The position on disc of the different targets are indicated by $\mu=\cos \theta$, where $\theta$ is the heliocentric angle. Blank spaces indicate that the data is not available. MPIL stands for magnetic polarity inversion line. $^{*}$In \citet{Stenflo2000}, the polarization signals are normalized by the intensity spectrum, i.e. $Q(\lambda)/I(\lambda)$ (\%).} \label{Signals}     
\end{table}

\begin{figure*}
\begin{center} 
 \includegraphics[trim=0 0 0 0,width=17.0cm]{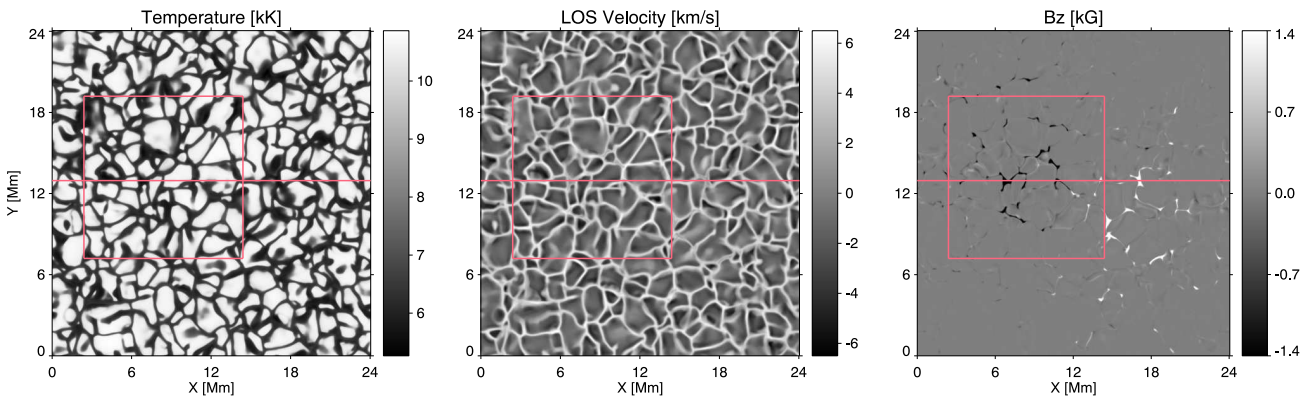}
 \vspace{-0.2cm}
 \caption{Snapshot 385 from the {\sc bifrost} enhanced network simulation. From left to right, temperature, LOS velocity, and longitudinal field strength at the geometrical height $Z=0$~km. We used two regions in this work, the one enclosed inside the squared box and the one highlighted with a horizontal line.}
 \label{context}
 \end{center}
\end{figure*}

For understanding why the Na {\sc i} $D$-lines produce larger scattering polarization signals than the Mg {\sc i} $b$ lines, we refer the reader to the paper by \cite{TrujilloBueno2001} on the Mg {\sc i} $b$ lines and to the works  of \cite{Belluzzi2013} and \cite{Belluzzi2015} on the Na {\sc i} $D$-lines. Regarding the linear polarization signals observed by \cite{Stenflo2000} in the Mg {\sc i} $b_1$ and $b_2$ lines, it was shown by \cite{TrujilloBueno2001} that they are due to the selective absorption processes produced by the presence of atomic polarization in their (metastable) lower levels \citep[see also,][]{TrujilloBueno2009}. In contrast, as shown by \cite{Belluzzi2013}, lower-level polarization is not needed to explain the scattering polarization observed in the Na {\sc i} $D$-lines \citep[see also,][]{Belluzzi2015}.

Among recent publications presenting observations of the Mg~{\sc i}~$b$ lines, we have works examining the Stokes profile properties for sunspot observations \citep{Deng2010}, the reversed granulation \citep{Rutten2011} or the visibility of Ellerman Bombs \citep{Rutten2015}. Moreover, there are works that used this line to connect the photosphere to upper layers, tracing the rise of quiet Sun magnetic loops and how much energy they carry with them \citep{MartinezGonzalez2009}. Finally, we have that, in all the cited cases, the Mg~{\sc i}~$b$ lines provide a unique opportunity for understanding the upper photosphere-low chromosphere, at heights that are in between those covered by traditional photospheric and chromospheric lines.

\subsection{Synthesis of the Stokes profiles}

We make use of the {\sc rh} code \citep{Uitenbroek2001,Uitenbroek2003} to synthesise the Stokes profiles. This code is able to compute atomic populations of the levels involved in the transitions under non-local thermodynamic equilibrium (NLTE) conditions and includes polarization resulting from the Zeeman effect. In particular, we assume the so-called field-free approximation \cite[][]{Rees1969}, i.e. we ignore the atomic level polarization and the fact that the radiation field that induces the atomic transitions is actually polarized \citep[see][for more details]{Bruls1996,TrujilloBueno1996}. This means that the main target of this work is to examine the Zeeman polarization signals, dominant in regions of strong magnetic fields, e.g. network patches, pores, or sunspots.

We use two types of input atmospheric models. First, we start with the semi-empirical FALC atmosphere \citep{Fontenla1993}. Later, we use  snapshot 385 of the enhanced network simulation \citep[][]{Carlsson2016} computed with the {\sc bifrost} code \citep{Gudiksen2011}. We show the spatial distribution of several atmospheric parameters at a geometrical height $Z=0$~km for the computed full field of view in Figure~\ref{context}. The horizontal line and the squared box designate the regions we use later on. Regarding the properties of this simulation, we refer the reader to the work of \cite{Carlsson2016}. We do not include any spatial degradation in our studies, i.e. we use the original horizontal pixel size of 48~km. 

Previous works noted that there is a lack of small scale random motions in this simulation \citep{Leenaarts2009} leading to narrower profiles than those in solar observations \citep[see also][]{delaCruzRodriguez2012} and for this reason we add a microturbulence of 1.5~km/s constant with height. This value is estimated from the microturbulence of the FAL models at around 750~km \citep[e.g,][]{Fontenla1990}. We use a spectral sampling of 10~m\AA \ and no spectral degradation is considered in this work. This means that the polarization signals we discuss later are slightly larger than what we expect from observations. In addition, we always normalize the synthetic profiles using the intensity of the local continuum spectrum. All the computations are done assuming disc centre observations, i.e.  $\mu=1$, where $\mu=\cos(\theta)$ and $\theta$ the heliocentric angle. Concerning the abundance of the different atomic species, we use the values provided in \cite{Asplund2009}. We take into account the collisional line broadening due to collisions with neutral hydrogen following  the theory of \cite{Anstee1995,Barklem1998}. The $\log$~$gf$ values of the Mg~{\sc i}~$b$ transitions presented in Table~\ref{lines} are computed directly from the atomic model, while those for the rest of the spectral lines are extracted from the database of R.~Kurucz \citep{Kurucz1995}. Finally, the Land\'{e} factors presented in Table~\ref{lines} are calculated from the principal quantum numbers, assuming LS coupling.

\subsection{Atomic data}

We consider three model atoms for calculating line profiles of the Mg~{\sc i} $b$ lines, representing different balances between realism and numerical expediency. Our aim is to find a model with minimal numbers of levels and transitions that can be employed to model Mg~{\sc i}~$b$ spectra in computationally demanding situations such as those resulting from 3D MHD simulations \citep[e.g., ][]{Leenaarts2009,Stepan2015,Stepan2016,Sukhorukov2017,Bjorgen2017}, or NLTE inversions of the Stokes spectra \citep[for instance,][]{SocasNavarro2015,delaCruzRodriguez2016}, where a simplified atom has a large impact in the total required computational time \citep[see also][]{Leenaarts2010}, while still retaining sufficient realism.

While the Mg~{\sc i} $b$ lines arise from the metastable $\rm 3s3p$~$\rm {}^3P$ levels, and their source functions are thus mostly set by the transfer in the lines themselves, their opacity strongly depends on the ionization equilibrium between the neutral stage and the first ionized stage. In the solar atmosphere most magnesium (90\% or more) occurs in the latter stage \citep{Carlsson1992}, making the population numbers in the neutral stage vary strongly with relatively small changes in the ionized stage. The NLTE ionization balance is established by the competing processes of radiative over-ionization in the UV and collisional recombination flow through high-lying (in the sense of energy levels) Rydberg states \citep[e.g,][]{Carlsson1992}. It follows that both processes need to be represented realistically in the model atom in order to establish a proper ionization balance and accurate population numbers in the neutral state.

To accurately estimate the UV radiative over-ionization (caused by the ionizing radiation field being hotter than the local temperature in the $b$-lines forming region of the atmosphere) we need to represent the UV line haze properly. Instead of including many spectral lines in the UV, which would require many wavelength points in the calculation of the Mg~{\sc i} bound-free continua, we apply a so-called Opacity Fudge (OF) according to the recipe described in \cite{Bruls1992}, applying frequency-dependent multiplication factors to the total opacity of H$^{-}$ for $208<\lambda<420$~nm and to the metal opacity at shorter wavelengths, i.e. $150<\lambda<210$~nm. These factors were determined empirically by fitting the computed VAL3-C \citep{Vernazza1981} continuum intensity to the observed continuum in such a way that the photospheric opacity source with the largest relative contribution is the one that is scaled at each wavelength. With the increased opacities the ionizing radiation field for magnesium is closer to the Planck function at the local temperature, reducing over-ionization and increasing population numbers in the neutral stage over the case without opacity corrections.

To properly represent the recombination flow coming from the near LTE populations of the Mg~{\sc ii} ground state requires the inclusion of many higher lying neutral magnesium levels and transitions between them at the cost of significantly increased computational effort, and increased risk of numerical instabilities. If these higher energy states are omitted, recombination is reduced and neutral magnesium populations are further below their LTE values, reducing opacity in the $b$ lines, and reducing the formation height of the lines. Therefore, the lines become weaker in this case.

\begin{figure}
\begin{center} 
 \includegraphics[trim=0 0 0 0,width=8.3cm]{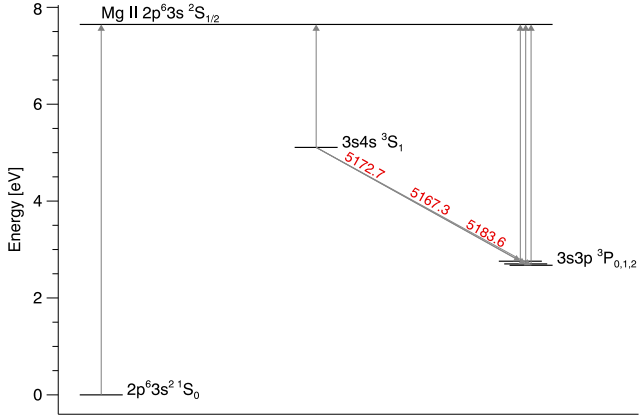}
 \vspace{-0.10cm}
 \caption{Grotrian diagram of the simplified Mg~{\sc i} 6 level atom used in this work.}
 \label{Grotrian}
 \end{center}
\end{figure}

To illustrate the above  effects we experimented with three atomic models. The first is the 68 level, 317 line and 67 bound-free transitions reference atom used by \citet{Carlsson1992} to realistically model the Mg~{\sc i} 12 $\mu$m lines (but here with the lower term of the $b$-lines split into the individual 3 levels). It includes many levels high in the Grotrian diagram up to $n = 10$, and the transitions between them. Combined with the OF recipe implemented in the {\sc rh} code this model reproduces the $b$-line profiles of the average quiet Sun very well using the 1D hydrostatic model C of \citet{Fontenla1993}. The other two, reduced size, atomic models are described below.

Our first alternative is an intermediate-size atomic model with 13 levels, see Table~\ref{levels12}, and 44 line transitions. We construct this atom from the 68 level model atom by replacing individual levels of similar properties with superlevels, using a procedure similar to that used in  \citet{2008ApJ...682.1376B}. We merge levels with similar effective quantum numbers $n^*$ thus creating six superlevels for $n^*=4,5,6,7,8,9$. We also merge the levels $\rm 4p\,^3P^{o}$, $\rm 3d\,^3D$ and $\rm 4p\,^1P^{o}$ into one superlevel and we neglect the intermediate singlet levels that are not important for the ionization balance. This atom is similar in size as the one presented in \cite{Mauas1988} but our use of superlevels for highly excited levels preserve the recombination flow that is important for the ionization balance.   Besides using a different way to craft the atom, we also treat the triplet $\rm 3s3p$~$\rm {}^3P^{o}$ term as three individual levels instead of as a single level.  In contrast to \cite{Mauas1988} we can thus generate the three different Mg~{\sc i}~$b$ spectral lines separately, which is important for this work as we aim to study the properties of the three Mg~{\sc i}~$b$ lines.

\begin{figure}
\begin{center} 
 \includegraphics[trim=0 0 0 0,width=7.2cm]{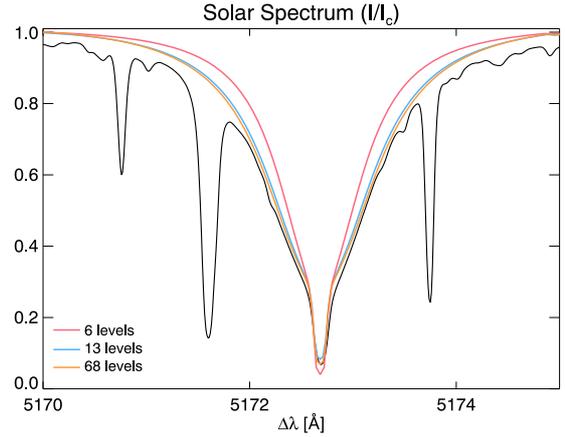}
 \vspace{-0.2cm}
 \caption{Comparison between the synthetic Mg~{\sc i}~$b_2$ Stokes $I$ profile computed using the 6 (red), 13 (blue), and the 68 (orange) level atom. We employ the FALC model as input atmosphere and the black line corresponds to the atlas.}
 \label{5vs12level}
 \end{center}
\end{figure}

The second one is a very simplified model, of just 6 levels and 3 line transitions. A detailed description of those levels, that we extracted from the {\sc nist} data base \citep{nist2015}, can be found in \cite{Martin1980} and also in Table~\ref{levels}. We also show a Grotrian diagram of the 6 level model in Figure~\ref{Grotrian}. The atom contains only three bound-bound transitions, those corresponding to the Mg~{\sc i} $b$ lines, between the level $\rm 3s4s$~$\rm {}^3S$ and the triplet  $\rm 3s3p$~$\rm {}^3P_{0,1,2}^{o}$. Collisional cross sections for the Mg~{\sc i} transitions in the smaller models are computed using the Seaton impact parameter approximation \citep[see][for more information]{Seaton1962}, collisional data are obtained from \cite{Sigut1995} and the collisional ionization from \cite{Allen1964} (see Pag. 42). Photo-ionization cross sections are extracted from the TOPbase atomic data base \citep{Cunto1992} for all transitions.

\begin{center}
\begin{table}
\hspace{-0.6cm}
\small
\begin{adjustbox}{width=0.42\textwidth}
  \bgroup
\def\arraystretch{1.25}
\begin{tabular}{lccccccc}
	\hline
Level & 	Energy [cm$^{-1}$]    & Configuration & Term  & g  \\
	\hline
0      &   0.0000000         &  $\rm 2p^{6}3s^{2}$ & $\rm {}^1S$      & 1    \\ 
1      &   21850.405        &  $\rm 2p^{6}3s3p$   & $\rm {}^3P^{o}$  & 1    \\
2      &   21870.464         &  $\rm 2p^{6}3s3p$   & $\rm {}^3P^{o}$  & 3    \\
3      &   21911.178         &  $\rm 2p^{6}3s3p$   & $\rm {}^3P^{o}$  & 5    \\
4      &   41197.403        &  $\rm 2p^{6}3s4s$   & $\rm {}^3S$      & 3    \\
5      &   48075.038         &  $\rm merged$   & $\rm {}$  & 27    \\
6      &   54219.683        &  $\rm n^*=4$   & $\rm {}$  & 64    \\
7      &   57079.214         &  $\rm n^*=5$   & $\rm {}$  & 100    \\
8      &   58539.650         &  $\rm n^*=6$   & $\rm {}$  & 144    \\
9     &   59392.276         &  $\rm n^*=7$   & $\rm {}$  & 189    \\
10     &   59936.323         &  $\rm n^*=8$   & $\rm {}$  & 249    \\
11     &   60306.014         &  $\rm n^*=9$   & $\rm {}$  & 308    \\
12     &   61671.020         &  $\rm 2p^{6}3s$     & $\rm {}^2S$      & 2    \\
	\hline
  \end{tabular}
  \egroup
\end{adjustbox}
\caption{Atomic levels for the Mg~{\sc i} 13 level atom used for comparison in this work. Each column, from left to right, contains the label assigned to each level, the energy, the spectroscopic notation, and the statistical weight.  Levels with different energy but identical term and configuration correspond to different values of total angular moment $J$. Level 5 is a superlevel merging $\rm 4p\,^3P^{o}$, $\rm 3d\,^3D$ and $\rm 4p\,^1P^{o}$. Levels 6-11 are superlevels merging levels with a given effective quantum number $n^*$. Level 12 corresponds to the first singly ionized level (the Mg~{\sc ii} ground term).}\label{levels12}     
\end{table}
\end{center}

\begin{figure*}
\begin{center} 
 \includegraphics[trim=0 0 0 0,width=16.5cm]{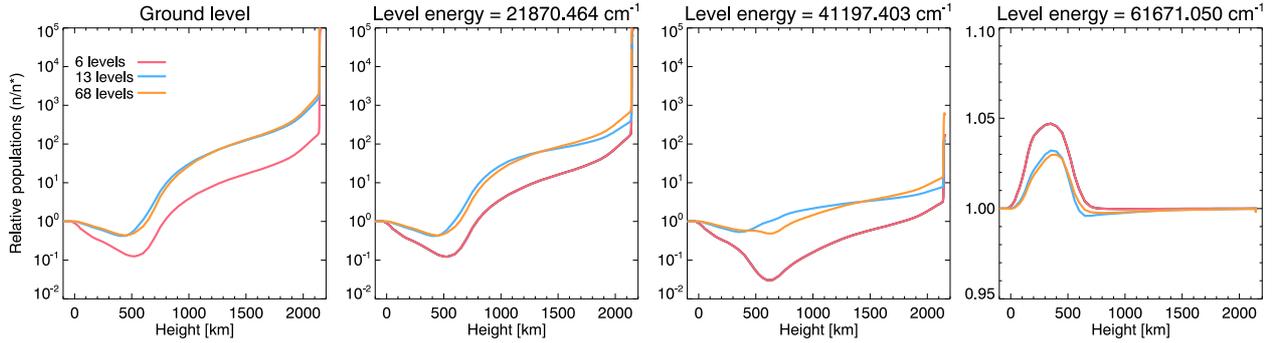}
 \vspace{-0.1cm}
  \caption{Comparison between the atomic populations of the ground level (leftmost panel), the two levels involved in the Mg~{\sc i}~$b_2$ transition, and the Mg~{\sc ii} ground level (rightmost panel). We represent the ratio between the atomic populations computed considering NLTE ($n$) and those obtained assuming LTE ($n^{*}$), i.e. the departure coefficients. Red designates the results from the simplified atomic model while blue and orange correspond to the 13 and 68 level atoms, respectively.}
 \label{Populations}
 \end{center}
\end{figure*}

\begin{figure*}
\begin{center} 
 \includegraphics[trim=0 0 0 0,width=16.5cm]{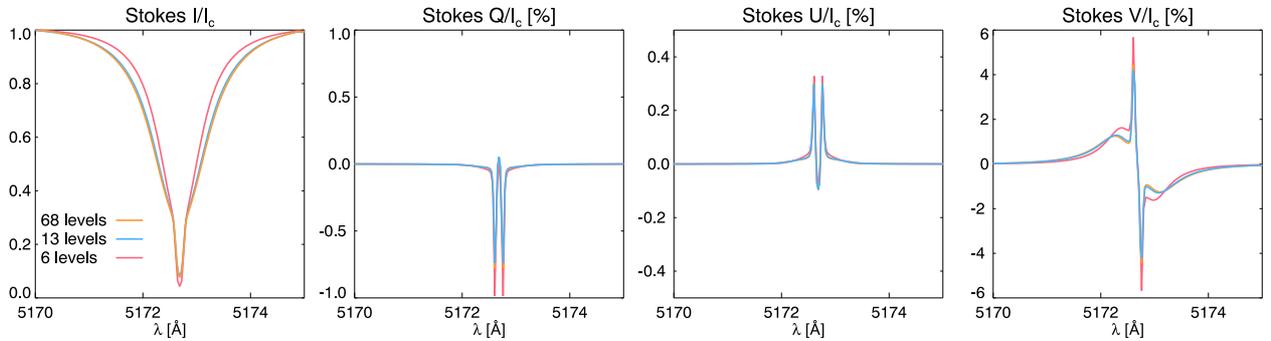}
 \vspace{-0.1cm}
  \caption{Comparison of the Stokes profiles generated by the three different model atoms using the FALC atmosphere with a field strength of 1000~G, 45 degrees of inclination, and 70 degrees of azimuth (constant with height).  We follow the colour code used in Figure~\ref{5vs12level} and we display, from left to right, Stokes ($I$, $Q$, $U$, $V$).}
 \label{Dif6_13_68levels}
 \end{center}
\end{figure*}

\begin{center}
\begin{table}
\hspace{-0.6cm}
\small
\begin{adjustbox}{width=0.42\textwidth}
  \bgroup
\def\arraystretch{1.25}
\begin{tabular}{lccccccc}
	\hline
Level & 	Energy [cm$^{-1}$]    & Configuration & Term  & g  \\
	\hline
0      &   0.00000000         &  $\rm 2p^{6}3s^{2}$ & $\rm {}^1S$      & 1    \\ 
1      &   21850.4050         &  $\rm 2p^{6}3s3p$   & $\rm {}^3P^{o}$  & 1    \\
2      &   21870.4640         &  $\rm 2p^{6}3s3p$   & $\rm {}^3P^{o}$  & 3    \\
3      &   21911.1780         &  $\rm 2p^{6}3s3p$   & $\rm {}^3P^{o}$  & 5    \\
4      &   41197.4030         &  $\rm 2p^{6}3s4s$   & $\rm {}^3S$      & 3    \\
5     &   61671.0200         &  $\rm 2p^{6}3s$     & $\rm {}^2S$      & 2    \\
	\hline
  \end{tabular}
  \egroup
\end{adjustbox}
\caption {Atomic levels used for crafting the simple Mg~{\sc i} 6 level atom. The table format is the same as in Table~\ref{levels12}.}\label{levels}     
\end{table}
\end{center}

\begin{center}
\begin{table}
\hspace{-0.6cm}
\small
\begin{adjustbox}{width=0.51\textwidth}
  \bgroup
\def\arraystretch{1.25}
\begin{tabular}{lccccc}
	\hline
 &  $I_{\rm core}/I_c [\%]$ &  $Q_{\rm max}/I_c [\%]$    &  $U_{\rm max}/I_c [\%]$ &  $V_{\rm max}/I_c [\%]$    \\
	\hline
6 level      &      43.7      &  26.5 & 8.8      & 27.5    \\ 
13 level     &    11.9        &  4.9   & 0.96  & 5.6    \\
	\hline
  \end{tabular}
  \egroup
\end{adjustbox}
\caption{Relative maximum difference between the results from the 68 level atom (used as reference) and the 6 (top) and 13 (bottom) level atoms. Columns show, from left to right, the differences between the line core intensity, maximum $Q$, $U$, and $V$ signals. See also Figure~\ref{Dif6_13_68levels}.}\label{diffmodels}     
\end{table}
\end{center}

\subsubsection{Comparions between different Mg~{\sc i} atoms}

We first study the differences between the intensity profiles generated by the 68, 13 and 6 level atoms. We display in Figure~\ref{5vs12level} the comparison between the synthetic profiles generated using those atoms and the solar atlas for the Mg~{\sc i}~$b_2$ line (we reduce the spectral range in this section compared to that showed in Figure~\ref{atlas} in order to visualise in detail the differences between atoms). We can see that the most complex atomic model (68 levels) matches the spectral line width and line core intensity. This is also true for the simplified 13 level atom with almost the same level of accuracy. In the case of the 6 level atom, we see larger differences with a profile that is narrower than that of the solar atlas.

\begin{figure*}
\begin{center} 
 \includegraphics[trim=0 0 0 0,width=15.0cm]{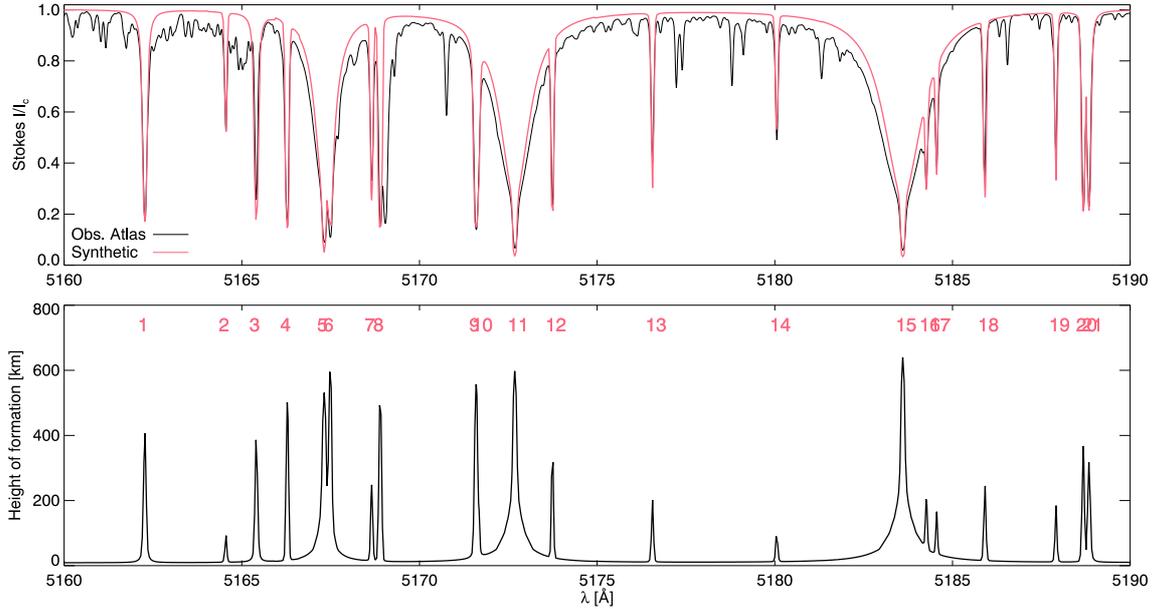}
 \vspace{-0.2cm}
 \caption{Upper panel shows the comparison between the observed atlas (black) and synthetic intensity spectrum (red) while the bottom panel displays the height where the optical depth is unity using as input atmosphere the FALC model.}
 \label{height_falc}
 \end{center}
\end{figure*}

\begin{figure*}
\begin{center} 
 \includegraphics[trim=0 0 0 0,width=15.5cm]{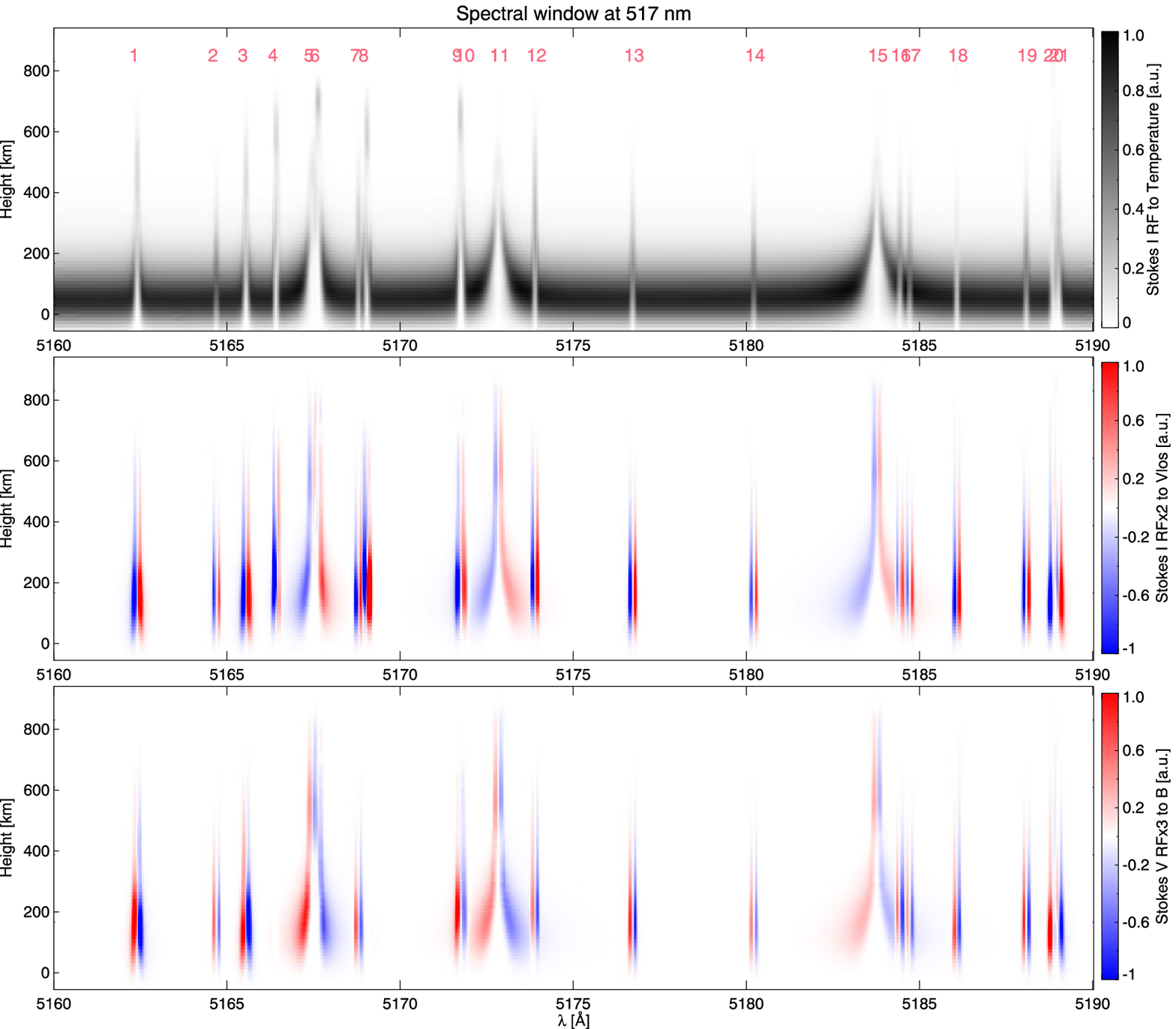}
 \vspace{-0.1cm}
 \caption{Response functions to changes in temperature (top), LOS velocity (middle), and field strength (bottom). White colour indicates no sensitivity to changes in the atmospheric parameters, while colour (or black) means that the spectral lines are sensitive to a perturbation of a given atmospheric parameter at a given height. We use the FALC atmospheric model and all the cases are normalized to the maximum of the Stokes $I$ RF to changes in the temperature (top panel).}
 \label{RF}
 \end{center}
\end{figure*}

Secondly, we examine the NLTE population departures (the ratio between the atomic populations computed in NLTE over those obtained assuming LTE) for the levels involved in the Mg~{\sc i}~$b_2$ transition. Starting with the ground level (leftmost panel of Fig.~\ref{Populations}) we can see that the 68 and 13 level atoms show similar populations at all heights, with the latter displaying slightly larger populations at middle heights. In the case of the 6 level atom, its ground level is underpopulated in comparison with the other two models. This is also true for the atomic levels that generate the Mg~{\sc i}~$b_2$ transition. Increasing the number of higher energy levels from the 6 to 13 to 68 level-models clearly reduces the NLTE underpopulation as described above. In addition, the upper and lower levels of the $b_2$ line behave very similarly, indicating that the source function of the line (proportional to the ratio of the upper to lower level population) is not affected by the number of atomic levels, in contrast to the opacity (which is proportional to the lower level population). In the case of the ground state of ionized magnesium (rightmost panel of Figure~\ref{Populations}), the population is almost the same in NLTE as in LTE and the simplified 6 level atom leads to a slight overpopulation.

 We also compare the polarization Stokes profiles using the three different atoms  (Figure~\ref{Dif6_13_68levels} and Table~\ref{diffmodels}). We find deviations, for instance in Stokes~$V$, between the simple atom and the other two models. In particular, we can see that, in the case of the 6 level atom, those differences  can be as high as 27.5\% of the Stokes~$V$ maximum amplitude (although in all the cases the shape of the profile is identical). Therefore, we must recommend caution when using the simplified atom. It is true that it is faster and simpler than the other two atoms, but if the aim is to achieve high accuracy,  the 13 level atom is recommended.

Finally, in terms of computational time, synthesizing the Stokes profiles shown in Figure~\ref{Dif6_13_68levels} took 2, 14, and  441 seconds for the 6, 13, and 68 level atoms, respectively. Thus, there is a large computational time reduction between the two simplified models and the most complex one. Moreover, although the difference between the 6 and 13 level time seems not very large, the former is a factor 7 faster, something that would have a noticeable impact when performing NLTE inversions of a large number of Stokes profiles or 3D synthesis. Therefore, we let the user decide which atom they prefer to use although a reasonable approach is to start with the 6 level atom performing tests because it is fast and robust and then switch to the 13 level as it is more accurate. In our case, for this first study of the Mg~{\sc i}~$b$ lines, we use the 6 level atom for the reasons stated above.

\subsection{Partial redistribution and NLTE effects}

We start performing a similar study as in \cite{QuinteroNoda2017b} checking the importance of partial redistribution (PRD) and NLTE on the Mg~{\sc i}~$b_2$ spectral line. Our results are similar to those presented in the mentioned work: PRD effects are negligible for the intensity profiles even for low $\mu$ values and, hence, we plan to assume complete redistribution from now on when synthesizing the Mg~{\sc i}~$b$ lines. Concerning the impact of NLTE on the synthetic spectrum, we found that LTE profiles differ from those computed assuming NLTE for both the intensity and the other Stokes parameters. Therefore, NLTE effects should be considered when analysing the Mg~{\sc i}~$b$ lines. We refer the reader to Section 3 of \cite{QuinteroNoda2017b} for more details.

\section{Results}

\subsection{Semi-empirical FALC atmosphere}

We start the analysis of the spectra at 517~nm computing the intensity profiles and the height where the optical depth is unity for all the lines included in the spectral region presented in Figure~\ref{atlas}. The results are displayed in Figure~\ref{height_falc} and we can see that the synthetic spectrum reproduces the photospheric lines and the line core intensity of the Mg~{\sc i}~$b$ lines. In the case of the wings of the latter spectral lines, there are discrepancies (in spite of using the opacity fudge correction) that, as we mentioned before, could be attenuated with a more complex model. We also see differences outside the spectral regions where the strongest lines are located, i.e. the synthetic spectrum shows larger intensity values caused by the neglect of various weak solar spectral lines. Concerning the height of formation of the computed spectral lines, the line cores of the Mg~{\sc i}~$b$ lines are formed around $500-600$~km height with the Mg~{\sc i}~$b_1$ (label 15) the one that extends highest in the atmosphere, followed by Mg~{\sc i}~$b_2$ (label 11) and, finally Mg~{\sc i}~$b_4$ (label 5). Interestingly, the core of the iron lines blended (labels 6, 9, and 10) with Mg~{\sc i}~$b_2$ and $b_4$ also reach very high in the photosphere, around 500~km in the FALC model. 

\subsubsection{Response Functions}

Figure \ref{RF} shows the response functions (RF) for the lines included in Table~\ref{lines} to changes in the temperature (top), line of sight (LOS) velocity (middle), and field strength (bottom). We followed the method explained in \cite{QuinteroNoda2016} to compute the NLTE numerical \citep[an analytical approach has recently been presented in][]{Milic2017,Milic2018} response functions using the FALC model with a magnetic field of 1000~G constant with height, while the inclination and azimuth are equal to 45 and 70 degrees, respectively. 

Starting with the RF to changes in temperature we can see that the Mg~{\sc i}~$b$ lines are sensitive only at low heights in comparison with the rest of the spectral lines \citep[see also Figure~1 of][]{Kneer1980}. Similar results are found for the K~{\sc i} D$_1$ \& D$_2$ lines \citep{QuinteroNoda2017b} and the Na~{\sc i} D$_1$ \& D$_2$ lines \citep{Eibe2001}. The reason for this behaviour was briefly explained in \cite{QuinteroNoda2017b} but we also refer the reader to the detailed discussion presented in \cite{Rutten2011}. If we focus on the RF to changes in the LOS velocity, we find a different behaviour, where the sensitivity of the Mg~{\sc i}~$b$ lines reaches much higher layers, around 800~km. The same happens for the field strength (bottom panel). Those heights are clearly higher than that showed by the rest of the photospheric lines. Thus, it seems that, although the sensitivity to the temperature mainly comes from the middle photosphere, for the rest of the atmospheric parameters the line core is sensitive to changes in the low chromosphere, around $700-900$~km. Finally, we can see that, unfortunately, the RF for the Mg~{\sc i}~$b_4$ line (label 5) is severely contaminated by the blended Fe~{\sc i} line (label 6) making the interpretation of their signals difficult.

\subsubsection{Polarization signals}

\begin{figure}
\begin{center} 
 \includegraphics[trim=0 0 0 0,width=7.0cm]{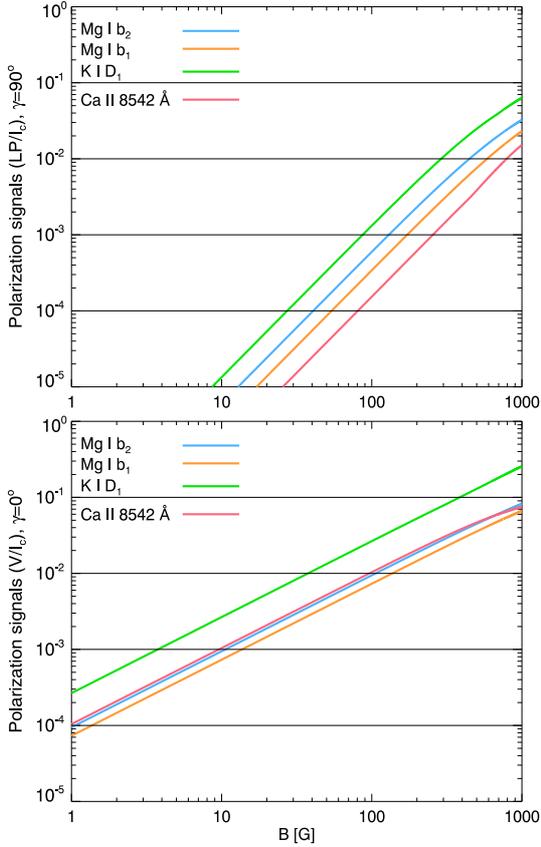}
 \vspace{-0.1cm}
 \caption{Maximum polarization signals for Mg~{\sc i}~$b_2$ (blue), Mg~{\sc i}~$b_1$ (orange), K~{\sc i}~$D_1$ (green), and Ca~{\sc ii} 8542~\AA \ (red). Upper panel shows the total linear polarization while the lower panel displays the maximum Stokes $V$ signals.}
 \label{Polsignals}
 \end{center}
\end{figure}

In order to compute the maximum polarization signals in a simple way, we first start with the FALC model used in the previous sections where we included a magnetic field constant with height. Then we modify its field strength using 1~G steps from 1 to 1000~G. As we plan to estimate the transversal and longitudinal signals separately, we perform the computation with a magnetic field inclination ($\gamma$) of 90~degrees first and, later with $\gamma=0$~degrees. In both cases, the magnetic field azimuth is fixed and equal to 70~degrees. In addition, we add the results for the K~{\sc i} D$_1$, and Ca~{\sc ii} 8542~\AA \ spectral lines \citep[presented in][]{QuinteroNoda2017, QuinteroNoda2017b} for comparison proposes. Moreover, we leave out from our studies from now on the Mg~{\sc i}~$b_4$ line  because it is not adequate for polarimetric observations as its polarization signals will be severely blended with those from the neighbouring photospheric line.

The top panel of Figure~\ref{Polsignals} displays the total linear polarization signals (computed as $LP=\sqrt{Q^2+U^2}$). We can see that the Mg~{\sc i}~$b$ lines (blue and orange) fall in between the K~{\sc i}~$D_1$ line that always show larger signals, and the Ca~{\sc ii} line, that is always weaker. In addition, we can also see that the Mg~{\sc i}~$b_2$ line (blue) shows stronger signals compared with the Mg~{\sc i}~$b_1$ (orange), something that we could predict as they are very similar lines, e.g. Doppler width, but the former has a larger Land\'{e} factor. Observing the Mg~{\sc i}~$b_2$ line,  we can expect polarization signals around $1\times10^{-3}$ of $I_c$ for a reference horizontal magnetic field of 100~G. 

Regarding the circular polarization signals (bottom panel of Figure~\ref{Polsignals}), we have again that the K~{\sc i}~$D_1$ shows larger signals than the rest of the studied lines. However, now we have that the Mg~{\sc i}~$b_1$ is weaker than the Ca~{\sc ii} line while Mg~{\sc i}~$b_2$ produces essentially the same signals. This is interesting in the sense that both, Mg~{\sc i}~$b_2$ and Ca~{\sc ii} 8542~\AA, are very broad lines (in comparison, for instance, with the K~{\sc i}~$D_1$ line), but the Mg~{\sc i}~$b_2$ line has a larger effective Land\'{e} factor ($g_{\rm eff}=1.75$ versus $g_{\rm eff}=1.10$). In order to understand this, we perform a thorough study of the circular and linear polarization sensitivity of these lines following the monograph of \cite{Landi2004}. The authors present a way of comparing the sensitivity of different spectral lines to the circular and linear polarization signals through the \textit{sensitivity index} \citep[see, for instance, the study presented in][]{MartinezGonzalez2008}. In this regard, linear and circular polarization signals scale linearly with those sensitivity indexes. The mentioned book explains in detail in chapter 9 all the implications of the different parameters that affect those sensitivity indices and apply them to different spectral lines. For the sake of clarity, we introduce here all the necessary equations to compute them (following the same notation used in the monograph) and we refer the reader to \cite{Landi2004} for more details.

\begin{table*}
\hspace{-0.9cm}
\normalsize
\begin{adjustbox}{width=0.90\textwidth}
  \bgroup
\def\arraystretch{1.25}
\begin{tabular}{lcccccccccccccc}
	\hline
Atom & $\lambda$ [\AA] &	$L_1$ & $L_2$  & $g_1$ & $g_2$ & $\bar{g}$ & $s$ & $d$ & $\bar{G}$  & $I(\lambda_0)$ [au]  & $d_c$ & $s_Q$ & $s_V$  \\
	\hline
Mg~{\sc i}   & 5172.68   & ${}^3P^{o}_{1}$   & ${}^3S_{1}$       &  1.50  &  2.00 & 1.75 & 4.00  & 0.00   & 2.88  & 666  & 0.933 & 1.700 & 2.877   \\		
Mg~{\sc i}   & 5183.60   & ${}^3P^{o}_{2}$   & ${}^3S_{1}$       &  1.50  &  2.00 & 1.25 & 8.00  & 4.00   & 1.53  & 592  & 0.940 & 1.219 & 1.547   \\
Fe~{\sc i}   & 5250.21   & ${}^5D_{0}$       & ${}^7D^{o}_{1}$   &  0.00  &  3.00 & 3.00 & 2.00  & -2.00  & 9.00  & 2841 & 0.713 & 2.255 & 7.104   \\
Fe~{\sc i}   & 6302.50   & ${}^5P^{o}_{1}$   & ${}^5D_{0}$       &  2.50  &  0.00 & 2.50 & 2.00  & 2.00   & 6.25  & 3444 & 0.652 & 2.066 & 6.510   \\
K~{\sc i}    & 7698.97   & ${}^2S_{1/2}$     & ${}^2P^{o}_{1/2}$ &  2.00  &  0.67 & 1.33 & 1.50  & 0.00   & 1.33  & 1652 & 0.833 & 1.710 & 2.632   \\
Ca~{\sc ii}  & 8542.09   & ${}^2D_{5/2}$     & ${}^2P^{o}_{3/2}$ &  1.20  &  1.33 & 1.10 & 12.50 & 5.00   & 1.21  & 1821 & 0.816 & 1.537 & 2.889   \\
	\hline
  \end{tabular}
  \egroup
\end{adjustbox}
\caption {Polarization sensitivity indices for a selection of spectral lines. From left to right we show the atomic species, line core wavelength, lower and upper level spectroscopic notation, Land\'{e} factor, effective \textit{first order} Land\'{e} factor, $s$ and $d$ (see Eq.~\ref{deq} and \ref{seq}), effective \textit{second order} Land\'{e} factor, line core intensity \citep[extracted from the solar atlas of][]{Delbouille1973}, $d_c$ (see Eq.~\ref{dceq}), and the circular and linear polarization sensitivity index. We define $\lambda_{\rm ref}$=5000~\AA \ and we assume $I_c=10000$~[au] for all the lines.} \label{index}     
\end{table*}

On one hand, the circular polarization sensitivity index is defined as

\begin{equation}
s_V=\left(\frac{\lambda_{0}}{\lambda_{\rm ref}}\right)\bar{g} \ d_c,
\label{circularindex}
\end{equation}
where $\lambda_{\rm ref}$ is an arbitrary reference wavelength, in our case $\lambda_{\rm ref}$=5000~\AA, and $\lambda_{0}$ the line core wavelength of the spectral line of interest. The effective Land\'{e} factor (assuming that all the lines fulfil the Russel-Saunders (or L-S) coupling scheme) is defined as 

\begin{equation}
\bar{g}=\frac{1}{2}(g_{1}+g_{2})+\frac{1}{4}(g_{1}-g_{2})[J_{1}(J_{1}+1)-J_{2}(J_{2}+1)],
\end{equation}
with $g_{i}$ and $J_{i}$ (with $i=1,2$) being the Land\'{e} factor, and the total angular momentum of the levels involved in the transition. This formula is invariant under interchange of the indices 1 and 2 \citep[see also][]{Shenstone1929}. The Land\'{e} factor of a given level is computed as 

\begin{equation}
g=\frac{3}{2}+\frac{S(S+1)-L(L+1)}{2J(J+1)},
\end{equation}
where $S$, $L$, and $J$, are the total spin, orbital angular momentum and total angular momentum of each level. The last term of Equation \ref{circularindex} is defined as 
\begin{equation}
d_c=\frac{I_c-I(\lambda_0)}{I_c},
\label{dceq}
\end{equation}
with $I_c$ the intensity of the continuum adjacent to the line and $I(\lambda_0)$ the line core intensity. On the other hand, the linear polarization sensitivity index is defined as

\begin{equation}
s_Q=\left(\frac{\lambda_0}{\lambda_{\rm ref}}\right)^2 \ \bar{G} \ d_c,
\end{equation}
where the main difference with respect to the circular polarization sensitivity index is that it depends on the square of the wavelength ratio and on the \textit{second order} effective Land\'{e} factor $\bar{G}$. This factor is given by the expression
\begin{equation}
\bar{G}=\bar{g}^2-\frac{1}{80}\left(g_1-g_2\right)^2\left(16s-7d^2-4\right),
\end{equation}
with
\begin{equation}
s=[J_1(J_1+1)+J_2(J_2+1)], 
\label{seq}
\end{equation}
and
\begin{equation}
d=[J_1(J_1+1)-J_2(J_2+1)].
\label{deq}
\end{equation}
In order to compute the sensitivity indices we use the atlas of \cite{Delbouille1973} as a reference, with $I_c=10000$~[au] for all the lines as the atlas continuum intensity is normalized to this value. The results for the lines displayed in Figure~\ref{Polsignals} are presented in Table~\ref{index}. We also include the sensitivity indices for the Fe~{\sc i} 5250.2~\AA \ and 6302.5~\AA \ lines to check that our computations are correct \citep[their sensitivity indices can be found in Tables~9.3 and 9.4 of][]{Landi2004}. If we compare the indices of the Mg~{\sc i}~$b_2$ line with those from the Ca~{\sc ii} 8542~\AA, we have that $s_Q$ is lower for the latter line. However, the $s_V$ indices are almost identical for the two lines explaining the behaviour found in Figure~\ref{Polsignals}. There are also additional effects that play a role in the circular polarization signals, as they are larger in lines of heavy elements compared with light ones \citep{Landi2004} what could compensate the larger Doppler width of the Ca~{\sc ii} 8542~\AA \ line compared with the Mg~{\sc i}~$b_2$ line. Finally, we also should not forget that the circular polarization signals scale as $\lambda_0^2$ what increases the Ca~{\sc ii} signals. We believe that some of these effects are responsible for the larger circular polarization signals of K~{\sc i}~$D_1$ even when $s_V$ is smaller than that of the Mg~{\sc i} and Ca~{\sc ii} lines, although probably the main reason is that the Doppler width of the K~{\sc i}~$D_1$ is much smaller than that of the other two lines. Finally, we can see that  the Mg~{\sc i}~$b_2$ reference maximum circular polarization signals that we could expect for a vertical magnetic field of 100~G are larger (one order of magnitude) than that obtained for the linear polarization signals and an inclined magnetic field.

\begin{figure*}
\begin{center} 
 \includegraphics[trim=0 0 0 0,width=17.0cm]{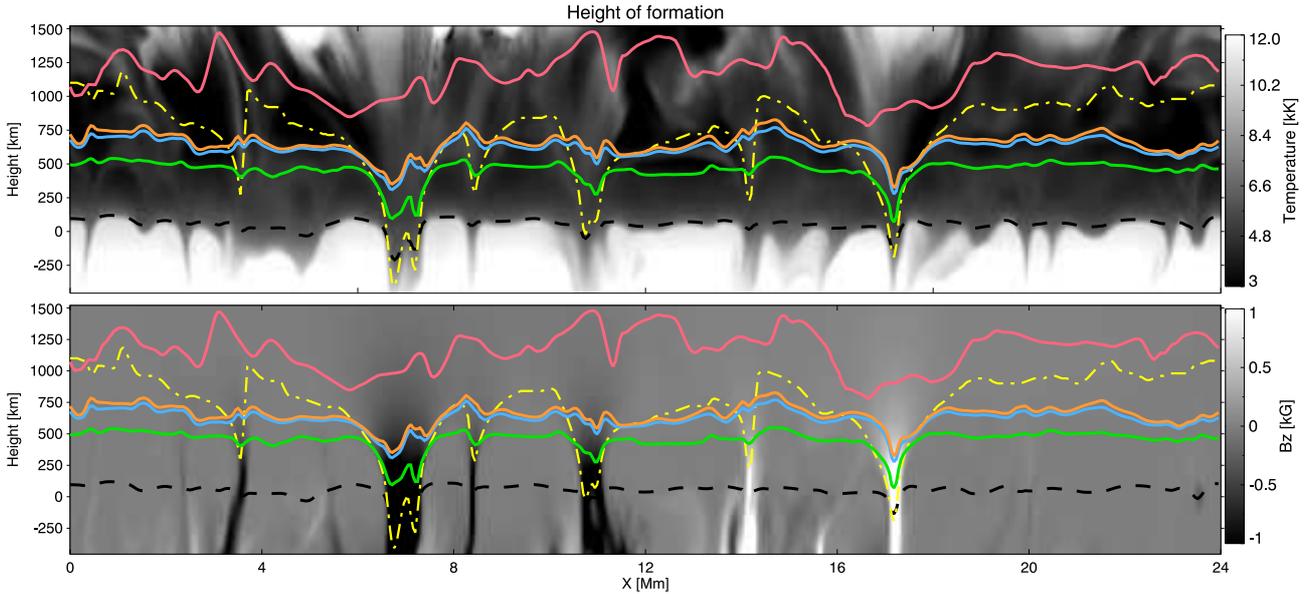}
 \vspace{-0.2cm}
 \caption{Height of formation of selected spectral lines. Each line designates the height where the optical depth is unity for the line core wavelength of Mg~{\sc i}~$b_2$ (blue), Mg~{\sc i}~$b_1$ (orange),  K~{\sc i} $\rm D_1$ (green), and Ca~{\sc ii} 8542~\AA \ (red). Background panels are the temperature (top) and longitudinal field (bottom) of the region indicated by the horizontal line in Figure~\ref{context}.  Dashed-dotted yellow line designates the region where the plasma $\beta\sim1$ and the dashed black line indicates the height where the optical depth is unity at 5000~\AA.}
 \label{height}
 \end{center}
\end{figure*}

\begin{figure*}
\begin{center} 
 \includegraphics[trim=0 0 0 0,width=17.0cm]{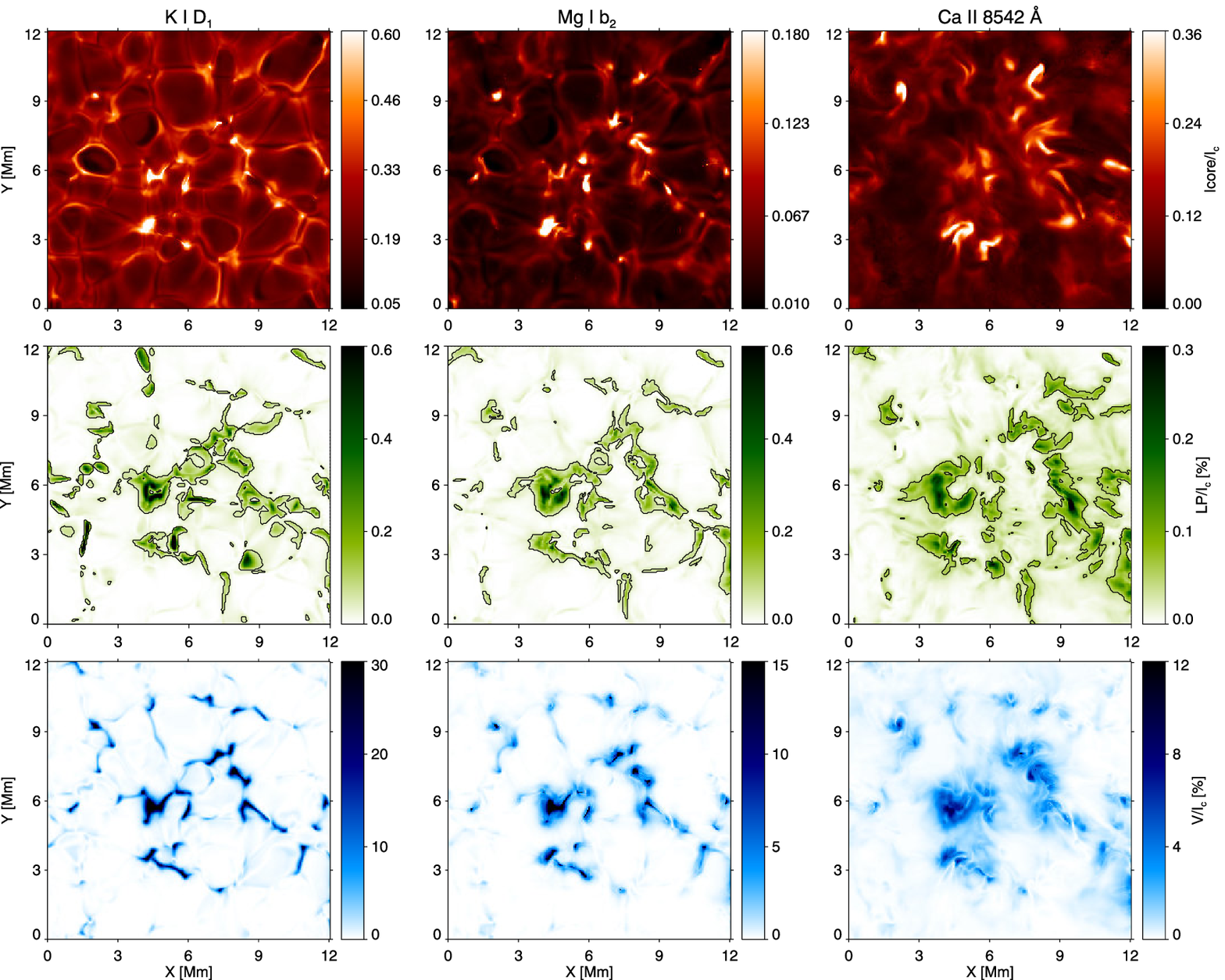}
 \vspace{-0.1cm}
 \caption{Spatial distribution of line core intensity (top), total linear polarization (middle), and circular polarization (bottom) for, from left to right, K~{\sc i}~$D_1$, Mg~{\sc i}~$b_2$, and Ca~{\sc ii}~8542~\AA, respectively. The displayed field-of-view corresponds to the highlighted square in Figure~\ref{context}. Contours on the linear polarization panels represent polarization signals larger than $5\times10^{-4}$ of $I_c$.}
 \label{Polsignals2D}
 \end{center}
\end{figure*}

\subsection{Bifrost enhanced network simulation}

\subsubsection{Height of formation}

We also compute the height where the optical depth is unity for the line core wavelength of the spectral lines analysed before using a slice (see the horizontal line in Figure~\ref{context}) of the snapshot 385 of the  {\sc bifrost} enhanced network simulation. We aim to understand how the heights of formation of the Mg~{\sc i}~$b_2$ and $b_1$ lines depend on different atmospheric parameters, in particular, on the magnetic field strength. In addition, we also include, as in the previous section, the the K~{\sc i} D$_1$, and Ca~{\sc ii} 8542~\AA \ lines results presented in \cite{QuinteroNoda2017b} for comparison proposes.

The results are displayed in Figure~\ref{height}, where we also include as a reference the height where the plasma $\beta$ (the ratio between the gas and the magnetic pressure) is unity (yellow) as well as where the optical depth is unity at 5000~\AA \ (black). The Mg~{\sc i}~$b$ lines form at very similar heights with Mg~{\sc i}~$b_1$ always slightly higher. Outside magnetic field concentrations the lines form below the $\beta=1$ region, around 750~km,  while in regions of strong magnetic fields the height of formation is similar or higher than that of  plasma $\beta=1$ and can drop up to 450~km.  In comparison with the other two lines, Mg~{\sc i}~$b$  always forms in between them, higher than the K~{\sc i}~$D_1$ (by about 200~km) and lower than the Ca~{\sc ii} 8542~\AA \ line. This indicates that combining, for instance, K~{\sc i}~$D_1$, Mg~{\sc i}~$b_2$, and Ca~{\sc ii} 8542~\AA \ lines, we would seamlessly cover the solar atmosphere from the upper photosphere to the middle chromosphere.

\subsubsection{Spatial distribution of polarization signals}

We complement the previous study computing the maximum polarization signals using the {\sc bifrost} simulation too. We focus on the region enclosed by the square in Figure~\ref{context} and we perform column by column computations, similar to what we did in previous works, using the one dimensional  geometry package of {\sc rh}. We leave for future studies a comparison between 1D and 3D computations as in \cite{Leenaarts2009,Stepan2016,Sukhorukov2017,Bjorgen2018} using the simplified model developed in this work. In addition, following the results from the previous sections, we can assume that observing the Mg~{\sc i}~$b_2$ and $b_1$ is not strictly necessary as they bring information from similar heights and Mg~{\sc i}~$b_2$ produces larger polarization signals for the same magnetic field configuration. Thus, in this section, in order to reduce the number of plotted maps, we just study the latter line, comparing its results with the results for the K~{\sc i}~$D_1$ and Ca~{\sc ii} 8542~\AA \ lines.

We show in Figure~\ref{Polsignals2D} the spatial distribution of the line core intensity (top), total linear polarization signals (middle), and maximum circular polarization signals (bottom) of the three mentioned spectral lines. Starting with the line core intensity, we can see the reverse granulation pattern in non-magnetic regions in the K~{\sc i}~$D_1$ line. This pattern is still present in some areas in the Mg~{\sc i}~$b_2$ line although much less defined, while in the rest of the selected field of view we find a chromospheric pattern dominated by threaded and complex structures that are rooted in the magnetic field concentrations. Interestingly, the spatial distribution of signals is different to those of the Ca~{\sc ii} line, something that it is because they are sensitive to different heights. Still, for the three lines we can see that magnetic field concentrations always enhance the line core intensity although producing different features for each spectral line.

If we examine the total linear polarization signals (middle row), we find that they are only present at the edges of magnetic field concentrations, being larger than $1\times10^{-3}$ of $I_c$ in the case of the flux concentration at (4,6)~Mm. For that case, we can see a clear pattern when we compare the three lines, where the magnetic field concentrations occupy wider areas as we examine the lines that form higher in the atmosphere. We are simply tracing the opening of the field lines due to the reduction of the gas pressure with height. This indicates that the three lines form and are sensitive to the atmospheric parameters at distinct atmospheric layers. The same effect can bee seen in the  circular polarization signals, where the clearly defined magnetic patches in the K~{\sc i}~$D_1$ line, fade and expand occupying larger areas for the rest of the lines. 

Finally, before moving to the summary of this work, we want to mention that if we examine the linear and circular polarization panels, we can see that there are not (or very weak) polarization signals beyond the location of the flux concentrations. This is because we are only taking into account the Zeeman effect during the computations. However, if additional mechanisms that produce polarization signals through scattering are included, we would expect larger signals in those weakly magnetized areas as was recently found by \cite{Stepan2016} for the Ca~{\sc ii} 8542~\AA \ spectral line. Unfortunately, estimating the amplitude of those signals is beyond the scope of this work.

\section{Summary}

We examined in this work the suitability of the Mg~{\sc i}~$b$ lines for chromospheric polarimetry. We revisited previous works available in the literature although we mainly focused on theoretical studies computing the Stokes profiles for different atmospheric conditions assuming the so-called field-free approximation and taking into account only the Zeeman effect. The first step we took was to develop two simplified atoms of 6 levels and 3 transitions and 13 levels and 44 transitions to reduce the computational time of the studies, enabling future applications like 3D synthesis or NLTE inversions. For this purpose, we make the constructed atom models publicly available.

Before studying the properties of the Mg~{\sc i}~$b$ lines we checked the capabilities of the simplified models comparing the synthetic profiles, as well as the populations of the different levels involved in the Mg~{\sc i}~$b$ transitions, with the results produced by a more complex Mg~{\sc i} atom of 68 levels and 317 transitions. We found that the 13 level atom contains almost the same information produced by the 68 level atom, what makes it ideal in terms of accuracy. On the other hand, the 6 level atom is extremely fast  (although less accurate) what makes it suitable for very demanding computational studies or for quick tests. Therefore, we make both models publicly available and we let the users decide which one is more suitable for their studies. 

In our case, we used for this work the 6 level atom for studying the three Mg~{\sc i}~$b$ lines separately and compare them. We only briefly examined the Mg~{\sc i}~$b_4$ line, because it is severely blended with Zeeman sensitive photospheric lines that complicates the interpretation of its polarimetric signals. Therefore, we do not consider it as a good candidate for polarimetric observations. After that, we focused on the two additional Mg~{\sc i}~$b$ lines. Our results show that Mg~{\sc i}~$b_1$ forms at slightly higher layers than those covered by Mg~{\sc i}~$b_2$. At the same time, the polarization signals in Mg~{\sc i}~$b_2$ are always larger. We explained that, as both lines are similar regarding the line core intensity and line width, this is simply due to the fact that the effective Land\'{e} factor of Mg~{\sc i}~$b_2$ is higher. Therefore, if the main target is to infer the magnetic properties of chromospheric structures, then the Mg~{\sc i}~$b_2$ is the most complete of the three Mg~{\sc i}~$b$ lines.

Finally, we compared the Mg~{\sc i}~$b_2$ line with additional upper photospheric/chromospheric lines, in particular, the K~{\sc i}~$D_1$ and Ca~{\sc ii} 8542~\AA, using the results from our previous publications. Our aim was to find whether the Mg~{\sc i}~$b_2$ line  can complement the other two spectral lines, simulating simultaneous multi-line observations that are going to be available in the near future thanks to missions such as Sunrise-3, DKIST or EST. We computed the average height of formation and the spatial distribution of polarization signals using the {\sc bifrost} 3D enhanced network simulation. Our results indicate that the Mg~{\sc i}~$b_2$ line is an excellent complement to the lines at 770~nm and 850~nm as it forms, and is sensitive to atmospheric parameters, at heights that are not covered by any of the lines included in the mentioned spectral windows. Therefore, the combination of these spectral lines will allow to seamlessly cover different atmospheric layers from the bottom of the photosphere to the middle chromosphere. This is crucial for the understanding, for instance, of the geometry of the magnetic field in different solar chromospheric features such as fibrils and mottles, or of how much energy is transported from the low photosphere to upper layers by magneto-hydrodynamic waves.

\section*{Acknowledgements}

We appreciate the help of the anonymous referee that, during the revision process, provided us comments and suggestions that allowed improving the manuscript. C. Quintero Noda acknowledges the support of the ISAS/JAXA International Top Young Fellowship (ITYF) and the JSPS KAKENHI Grant Number 18K13596. The SUNRISE-3 project is supported in Japan by the funding from ISAS/JAXA for the small-scale program for novel solar observations and the JSPS KAKENHI Grant Number 18H03723 and 18H05234. This research was supported by the Research Council of Norway through its Centres of Excellence scheme, project number 262622. This work has also been supported by Spanish Ministry of Economy and Competitiveness through the project ESP-2016-77548-C5-1-R.  D. Orozco Su\'{a}rez also acknowledges financial support through the Ram\'{o}n y Cajal fellowships. 

\bibliographystyle{mnras} 
\bibliography{magnesium} 

\bsp	
\label{lastpage}
\end{document}